\documentclass[oldversion]{aa}
\usepackage{txfonts}
\usepackage{graphicx}
\usepackage{natbib}
\bibpunct{(}{)}{;}{a}{}{,} 

\newcommand{\ergs}{\mbox{$\mathrm{erg\,s^{-1}}$}}
\newcommand{\ergcs}{\mbox{$\mathrm{erg\,cm^{-2}}\mathrm{s^{-1}}$}}
\newcommand{\ergcsa}{\mbox{$\mathrm{erg\,cm^{-2}}\mathrm{s^{-1}}\mathrm{\AA ^{-1}}$}}
\newcommand{\gcs}{\mbox{$\mathrm{g\,cm^{-2}}\mathrm{s^{-1}}$}}

\newcommand{\hsh}{\mbox{$h_\mathrm{sh}$}}
\newcommand{\Msun}{\mbox{$M_{\odot}$}}

\newcommand{\La}{\mbox{${\mathrm{Ly\alpha}}$}}

 \newlength{\width}

\begin{document}

\title{Irradiated atmospheres of accreting magnetic white dwarfs with
an application to the polar AM~Herculis}

\author{
M. K\"onig\inst{1} 
   \and
K. Beuermann\inst{1}
   \and 
B.T. G\"ansicke\inst{2}
}
 
\offprints{M. K\"onig, e-mail: mkoenig@astro. physik.uni-goettingen.de}

\institute{
Institut f\"ur Astrophysik der Universit\"at G\"ottingen, 
Friedrich-Hund-Platz 1, D-37077 G\"ottingen, Germany
  \and
Department of Physics, University of Warwick, Coventry CV4 7AL, UK
}

\date{Received October 11, 2005 ; accepted December 15, 2005 }

\abstract{We present a pilot study of atmospheres of accreting
magnetic white dwarfs irradiated by intense fluxes at ultraviolet to
infrared wavelengths. The model uses a standard LTE stellar
atmosphere code which is expanded by introducing an angle-dependent
external radiation source. The present results are obtained for an
external source with the spectral shape of a 10000\,K blackbody and a
freely adjustable spectral flux. The model provides an explanation
for the observed largely filled-up Lyman lines in the prototype polar
AM~Herculis during its high states. It also confirms the hypotheses
(i) that irradiation by cyclotron radiation and other radiation
sources is the principle cause for the large heated polar caps
surrounding the accretion spots on white dwarfs in polars and (ii)
that much of the reprocessed light appears in the far ultraviolet and
not in the soft X-ray regime as suggested in the original simple
theories. We also briefly discuss the role played by hard X-rays in
heating the polar cap.
\keywords{radiative transfer -- stars: atmospheres -- (stars:) novae,
cataclysmic variables -- (stars:) white dwarfs -- stars: individual
(AM Herculis)} }
   
\maketitle

\section{Introduction}
The white dwarfs in magnetic cataclysmic variables (mCVs, more
specifically polars) accrete in restricted regions near their magnetic
poles. In most of them, the infalling matter is heated in a
free-standing shock above the surface of the white dwarf and cools by
the emission of X-rays and cyclotron radiation. Early models assumed
that the stand-off distance of the shock is small and much of the
emission by the post-shock plasma is absorbed by the atmosphere below
and immediately around the accretion spot. The implied irradiation of
this small section of the atmosphere by intense fluxes of hard X-ray
bremsstrahlung and cylotron radiation leads to an expected temperature
of the heated atmosphere of $kT \simeq 25$~eV ($T \simeq 3\times
10^5$\,K), to re-emission in the soft X-ray regime with a
quasi-blackbody spectrum, and to an energy balance $L_\mathrm{sx} =
f\,[L_\mathrm{cyc}+(1-A_\mathrm{hx})\, L_\mathrm{hx}]$
\citep{kinglasota79,lambmasters79}, where the three luminosity
components refer to soft X-rays (sx), hard X-rays (hx), and cyclotron
radiation (cyc), the geometry factor $f\la1/2$ accounts for the
fraction of the luminosity intercepted by the white dwarf, and the
hard X-ray albedo $A_\mathrm{hx}$ reduces the efficiency of X-ray
heating \citep{vanteeselingetal94}. Later on, it was realized that the
accretion region is highly structured \citep{kuijperspringle82,
franketal88} with individual sections receiving vastly different mass
flow densities and displaying substantially different emission
characteristics. The plasma in tenuous tall shocks cools
preferentially by cyclotron radiation, the plasma in dense low-lying
shocks by hard X-ray bremsstrahlung, and for the highest mass flow
densities, the shocks are buried in the atmosphere and the primary
bremsstrahlung is reprocessed into soft X-rays, a component physically
different from but observationally difficult to distinguish from the
originally suggested soft X-ray blackbody component. This ``soft X-ray
puzzle'' was the subject of an extended debate over more than two
decades \citep[e.g.,][ and references therein]{beuermann04,
ramsaycropper04}.

Quite unexpectedly, observational support for the reprocessing
scenario came from studies of the far ultraviolet spectra of the
prototype polar AM~Herculis, which showed that photospheric emission
of the white dwarf dominates the FUV \citep{heiseverbunt88,
gaensickeetal95-1, gaensickeetal98-2, maucheraymond98, greeleyetal99,
gaensickeetal05-1}. The pronounced orbital modulation of the FUV flux
indicates a large heated polar cap, which covers $\sim 10$\% of the
white dwarf surface.  The cap reaches a peak temperature of $\sim
3.5\times 10^4$\,K in low states when accretion nearly ceases and
becomes much hotter in high states. \citet{gaensickeetal95-1}
demonstrated that the excess FUV flux of the polar cap quantitatively
agrees with the sum of cyclotron and X-ray fluxes from AM~Her, both in
the high and the low states, thus confirming the energy balance
predicted by the simple models, with the decisive difference, however,
that the heated polar cap covers a much larger area than the proper
accretion spot and the reprocessed flux emerges in the FUV and not at
soft X-ray wavelengths. AM~Her is the best observed polar, but FUV
studies of other polars suggest that the situation encountered in
AM~Her is quite typical of the class \citep[e.g.,][]{araujobetancoretal05-1}.

The dominant FUV spectral feature of white dwarfs in polars are the
Lyman absorption lines, which are deep and broad in low states,
mimicking the pure hydrogen spectra of DA white dwarfs, but become
filled up almost entirely in states of high accretion
\citep{heiseverbunt88, gaensickeetal95-1, gaensickeetal98-2,
maucheraymond98, greeleyetal99, gaensickeetal05-1}. \citet{gaensickeetal98-2} 
first tried to model the filled-up Lyman lines using a simple toy
model and concluded that the temperature stratification in the white
dwarf atmosphere is severely affected over the entire irradiated polar
cap.

A spot due to reprocessing covering 10\% of the white dwarf surface
requires a point-like source at a height of 0.25 white dwarf radii or
an extended source at a lower height. The best candidates are
optical/infrared cyclotron and to some extent hard X-ray emission from
tall shocks, supplemented by the FUV to optical emission from the
pre-shock accretion stream. A further potential source is shockless
accretion at very low mass flow densities, $\dot m <
10^{-3}$\,\gcs. The infalling particles do not directly heat the lower
atmosphere but create a hot corona and the atmosphere below is again
heated radiatively \citep{thomsoncawthorne87, woelkbeuermann92}. The
task is, therefore, to consider the properties of atmospheres heated
by a large external flux of ultraviolet/optical/infrared radiation.
Modeling the UV-emitting spots in polars requires us to consider
incoming radiation fluxes that exceed the flux emerging from the
unheated atmosphere of the white dwarf by factors up to $\sim200$ in
the high state. Irradiated stellar atmospheres have been studied by
several authors for a variety of conditions which differ, however,
from those considered here \citep[e.g.,][]{londonmccray81,
vanteeselingetal94, brettsmith93, barmanetal04}. In this paper, we
present a pilot study of the structure and the angle-dependent spectra
of white dwarfs irradiated by an intense source of infrared to
ultraviolet radiation.

The paper is arranged as follows. In Sect.\,2 we describe the stellar
atmosphere code used to calculate the irradiated atmosphere models,
Sect.\,3 introduces the essential parameters of the accretion regions
in polars, in Sect.\,4 we report the results for simple irradiation
geometries, which are applied to the prototype polar AM~Her in
Sect.\,5. In Sect.\,6, we discuss the limitations of the approach and
lines for future research.

\section{Radiation transfer in irradiated atmospheres}

We assume a pure hydrogen elemental composition, as appropriate for DA
white dwarfs and for the sections of the atmosphere of accreting white
dwarfs in mCVs that far away from the main accretion spot. We
introduce the external flux as a boundary condition and neglect any
other disturbance of the atmosphere as, e.g., an external pressure
exerted by low-level accretion outside the main accretion spot.

We use a standard LTE atmosphere code to solve the radiative transfer
in a plane parallel geometry with an incoming flux $I_\mathrm{in}(\mu,
\lambda)$ at optical depth $\tau=0$, prescribed as a function of
wavelength $\lambda$ and zenith angle $\vartheta$ via the direction
cosine $\mu =$\,cos$\vartheta$. The radiative transfer equation
\begin{equation}
\label{stg}
\mu \, \frac{\partial I(\tau, \mu, \lambda)}{\partial \tau}
= I(\tau, \mu, \lambda) - S(\tau, \lambda)
\end{equation}
accounts for absorption and isotropic electron scattering via the
source function 
\begin{equation}
\label{quellfunktion}
S = \frac{\kappa}{\kappa + \sigma} \, B +
\frac{\sigma}{\kappa + \sigma} \, J \;,
\end{equation}
where $\kappa$ is the absorption coefficient, $\sigma$ is the Thomson
scattering cross section, $B$ the Planck function, $J$ the
angle-averaged intensity, and the dependencies on $\tau$ and
$\lambda$ have been suppressed. Line opacity is included in the form
of the Lyman, Balmer, Paschen, and Brackett series of hydrogen,
including excited levels up to $n=5$. For the hot atmospheres
considered here, molecular absorption is unimportant. Although we
apply the results to magnetic white dwarfs in CVs we neglect Zeeman
splitting of the Lyman lines. This is appropriate for the moderate
field strengths considered here ($|\vec{B}| \la 30$\,MG), because
the Lyman lines are split by the normal Zeeman effect with $\Delta
\lambda=e\lambda^2|\vec{B}|/(4\pi m_{\rm e} c^2)$, where $e$ is the
elementary charge, $m_{\rm e}$ the electron mass. and $c$ the speed of
light. Specifically, for the polar AM~Her considered below, the
observed field strength $|\vec{B}|=14$\,MG implies $\Delta
\lambda=10$\AA\ for \La, which is small compared with the 90\AA\
contributed by the Stark effect to the width of \La\ at the white
dwarf effective temparature of 20000\,K.

The radiative transfer equation is solved by a standard \citet{rybicki71}
elimination scheme. To this end, the quantities
\begin{eqnarray}
\label{stggroessen}
u(\tau,\mu,\lambda) &=& 
\frac{1}{2} \left[ I(\tau,\mu,\lambda) + I(\tau,-\mu,\lambda) \right] \\
\upsilon(\tau,\mu,\lambda) &=& 
\frac{1}{2} \left[ I(\tau,\mu,\lambda) - I(\tau,-\mu,\lambda) \right]
\end{eqnarray}
are introduced for $0 \le \mu \le 1$. With their help,
Eq.~(\ref{stg}) is transformed to a second order differential
equation
\begin{equation}
\label{stg2}
\mu^2 \, \frac{\partial^2 u(\tau,\mu,\lambda)}{\partial \tau^2} =
u(\tau,\mu,\lambda) - S(\tau,\lambda)
\end{equation}
which is solved numerically. At large optical depths, the diffusion
approximation provides the standard boundary condition.  The
angle-dependent incident intensity $I_\mathrm{in}(\mu,\lambda)$ is
implemented by modifying the upper boundary condition
\begin{equation}
I(0,-\mu,\lambda) = I_\mathrm{in} (\mu,\lambda) \ne 0 \;.
\end{equation}
For isotropic irradiation, $I_\mathrm{in}$ is independent of $\mu$,
for irradiation at a single angle $\vartheta_0$ with
$\mu_0=$cos$\vartheta_0$, $I_\mathrm{in}$ is described by a
delta-function $\delta(\mu-\mu_0)$. The incoming and outgoing fluxes
at $\tau=0$ are denoted by
\begin{eqnarray}
F_\mathrm{in}(0,\lambda) & = &
2\pi\int_0^1\left[I(0,-\mu,\lambda)\right]\,\mu\,d\mu\;.\\
F_\mathrm{out}(0,\lambda) & = & 2\pi \int_0^1\left[I(0,\mu,\lambda)
\right]\,\mu\,d\mu.
\end{eqnarray}
Independent of the magnitude of the incoming flux, the integral
quantities
\begin{equation}
F_\mathrm{in} = \int_{\,0}^\infty
F_\mathrm{in}(\lambda)d\lambda\;~~~~{\rm and}~~~~ F_\mathrm{out} =
\int_{\,0}^\infty F_\mathrm{out}(\lambda)d\lambda
\end{equation}
are related by $F_\mathrm{out} = F_\mathrm{in} + F_\mathrm{wd}$, where 
\begin{equation}
F_\mathrm{wd} = \int_{\,0}^\infty F_\mathrm{wd}(\lambda)d\lambda =
\sigma T_\mathrm{wd}^4
\end{equation}
is the flux emerging from the unheated white dwarf atmosphere with
effective temperature $T_\mathrm{wd}$.

The condition of radiative equilibrium, 
\begin{equation}
\label{radequ}
\int_0^\infty \kappa \, \left( J - B \right) \, d\lambda = \int_0^\infty
\chi \frac{\partial H}{\partial \tau}\,d\lambda = 0\;,
\end{equation}
with $H=(1/4\pi)\,F$ the Eddington flux, is achieved by appropriate
depth-dependent adjustment of the temperature. We use a correction
method that is based on solving the momentum equations
\begin{equation}
\label{tempcorr1}
\frac{\partial^2 (f_K J)}{\partial \tau^2} = J - S, \qquad
\frac{\partial (f_K J)}{\partial \tau} = f_H J
\end{equation}
with Eq.~(\ref{radequ}) as a constraint.  Here, $f_H = H/J$ and $f_K =
K/J$ are the Eddington factors with $K=\frac{1}{2} \int_{-1}^1 I\,\mu^2\,d\mu$
the scalar radiation pressure.  A correction $\delta T$ to $T(\tau)$
is obtained by substituting B with
\begin{equation}
\tilde{B} = B[T(\tau) + \delta T(\tau)] 
\approx B(T) + \frac{d B(T)}{dT} \delta T(\tau)
\end{equation}
and solving Eq.~(\ref{tempcorr1}) simultaneously with
Eq.~(\ref{radequ}) for $\delta T$.

\section{Irradiated polar caps in AM~Her  stars}

% ------------  Fig. 1 --------------------------------------------------
\begin{figure}[t]
\vspace{1mm}
\includegraphics[angle=0,width=\columnwidth]{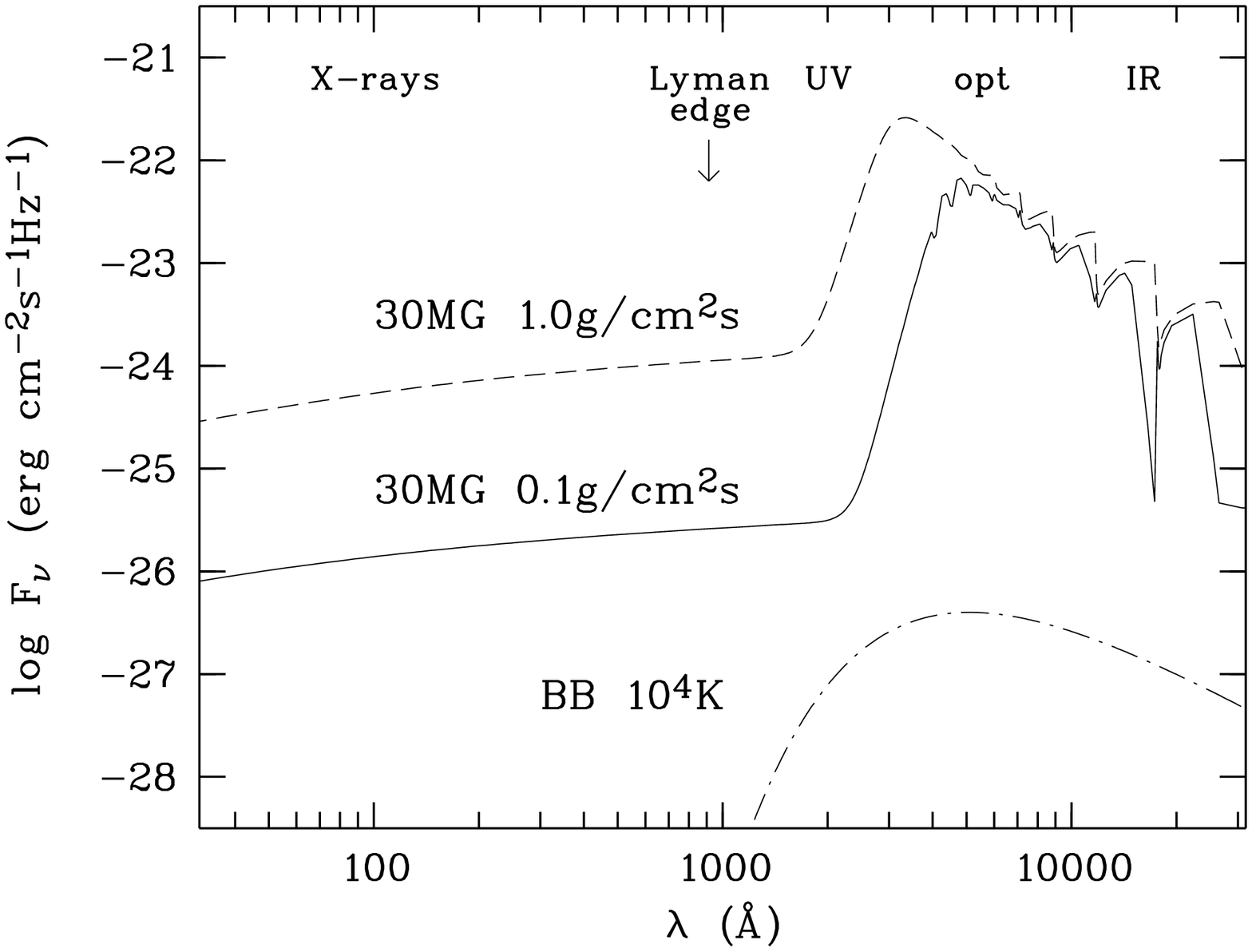}
\caption[ ]{Spectral energy distribution at an angle of $\theta =
80^\circ$ relative to the field direction for the post-shock
emission region on a 0.6\,\Msun\ white dwarf at a distance
$D=10$\,pc with field strength $|\vec{B}|=30$\,MG, area
$A=4\times10^{16}$\,cm$^2$, and mass flow densities of $\dot
m=1\,\gcs$ (dashed curve) and $0.1\,\gcs$ (solid curve) \citep[ their
Fig.\,9]{fischerbeuermann01}. For comparison, a blackbody with
temperature $T=10^4$\,K and the same emitting area is shown
(dot-dashed curve).}
\label{fig-1}
\end{figure}
% -------------------------------------------------------------------------

The frequency-integrated bremsstrahlung and cyclotron emissivities
vary as $\varepsilon_{\rm brems} \propto n_{\rm e}^2\,T_{\rm e}^{1/2}$
and $\varepsilon_{\rm cyc} \propto n_{\rm e}\,T_{\rm e}|\vec{B}|^2$,
respectively, with $n_{\rm e}$ the electron (and ion) number density
of the plasma, $T_{\rm e}$ its electron temperature, and $|\vec{B}|$
the field strength. The plasma temperature is of the order of
$10^8$\,K at the shock and the plasma cools at nearly constant
pressure ($n_{\rm e} \propto 1/T_{\rm e}$) as it moves down. Since
$|\vec{B}|$ is also approximately constant throughout the post-shock
region, the emissivities vary approximately as $\varepsilon_{\rm
brems} \propto T_{\rm e}^{-3/2}$ and $\varepsilon_{\rm cyc} \propto
T_{\rm e}^{0}$, i.e., bremsstrahlung emission is concentrated near the
white dwarf surface, while cyclotron emission originates throughout
the column and, on the average, high above the surface. For a
shock-heated plasma dominated by cyclotron cooling, the stand-off
distance of the shock is given by
\begin{equation}
h_{\rm sh} = 3.3\times10^8\,|\vec{B}|_7^{-2.6}M_1^{1.2}~{\rm cm},
\label{eq:shockheight}
\end{equation}
where $|\vec{B}|_7$ is the field strength in units of $10^7$\,G
and $M_1$ is the white dwarf mass in solar units \citep[ their
Eq.~21]{fischerbeuermann01}. Equation\,(\ref{eq:shockheight}) remains
valid as long as \hsh\ stays small compared with the radius of the
white dwarf. For a 0.6\,\Msun\ white dwarf with $|\vec{B}|=100$\,MG,
\hsh\ amounts to a minute $4\times10^5$\,cm, while for
$|\vec{B}|=10$\,MG, Eq.~(\ref{eq:shockheight}) yields \mbox{$h_{\rm
sh}\simeq1.8\times10^8$\,cm} or about 20\% of the white dwarf
radius. Large irradiated polar caps are, therefore, expected to exist
preferentially in low-field mCVs.

The angle-dependent spectra emitted by the accretion regions on
magnetic white dwarfs have been calculated by \citet[ see also Woelk
\& Beuermann, 1996]{fischerbeuermann01}. Figure~\ref{fig-1} shows an
example of the emitted spectral energy distribution in a direction
almost perpendicular to the magnetic field. Cyclotron emission is
largely optically thick in the infrared and assumes a quasi
Rayleigh-Jeans spectrum with harmonic structure superimposed. The
cyclotron spectrum breaks off in the optical/UV, where the emission
region becomes optically thin. At high frequencies, the X-ray
bremsstrahlung tail is seen, which contains negligible flux if the
mass flow density $\dot m$ is sufficiently low. This is a consequence
of the dependence of $\varepsilon_{\rm brems}$ on $n_{\rm e}^2 \propto
\dot m^2$. For comparison, we show a blackbody for a temperature of
$10^4$\,K with the same emitting area. Naturally, its spectral flux in
the infrared is lower by a factor of about $10^4$ compared with the
$10^8$\,K cyclotron source. The comparison shows that the general
spectral shape of the cyclotron emission mimics that of a blackbody of
about $10^4$\,K.  For the purpose of the present pilot study, the
exact shape and angle dependence of the cyclotron emission is not
critical. We, therefore, approximate the shape of the irradiation
spectrum by a $10^4$\,K blackbody and scale the flux to match that of
the cyclotron spectrum at the surface element in question. The reader
should realize that the irradiation spectrum belongs to the plasma
temperature of $\sim10^8$\,K and that the irradiation flux can,
therefore, substantially exceed that of a $10^4$\,K blackbody.

\section{Results}

All results presented in this paper have been obtained for irradiation
by a source that emits a spectrum with the shape of a blackbody
of $T_{\rm bb}=10^4$\,K.  Test calculations for various values of
$T_{\rm bb}$ between 3000\,K and 25000\,K suggest that the spectrum
emerging from the heated atmosphere is insensitive to $T_{\rm bb}$ as
long as the fraction of the emitted energy shortward of the Lyman edge
is small \citep{koenig04}.

In a first step, we consider single surface elements irradiated (1)
isotropically and (2) at a single zenith angle $\vartheta$. 
With respect to case 2, we note that our atmosphere code
treats the intensity $I(\tau,\vartheta)$ as azimuthally symmetric and,
hence, interpretes the incoming intensity at $\tau=0$ as given on a
cone with opening angle $\vartheta$ rather than at a single direction
$\vartheta,\varphi$ with azimuth angle $\varphi$. As a consequence,
the code neglects the azimuthal asymmetry in the outgoing intensity
that occurs in nature for a single surface element irradiated
unidirectionally from $\vartheta,\varphi$. In a second step, we use
case-2 models to calculate the integrated intensity from all surface
elements of the white dwarf visible from a given direction
$\theta,\phi$, which indicates the direction towards the observer. We
expect that the asymmetries present in the emissions from the
individual elements largely cancel in the integral over all surface
elements. With this caveat, we calculate the integrated spectrum
received from an entire white dwarf, which possesses a large hot spot
produced by either a point source or an extended source of total
luminosity $L$ located at a height $h$ above the photosphere.

We describe the wavelength-integrated irradiation flux $F_\mathrm{in}$
of Eq.~(9) in units of $F_\mathrm{wd}$ of Eq.~(10) by
\begin{equation}
F_\mathrm{in} = x~F_\mathrm{wd}
\end{equation}
and consider irradiation parameters $x$ up to 200. We use an unheated
white dwarf atmosphere with effective temperature
$T_\mathrm{wd}=2\times 10^4$\,K and log\,$g=8$, the
approximate parameters of the white dwarf in AM~Her
\citep{gaensickeetal95-1,gaensickeetal05-1}. The chosen effective
temperature is typical also of the white dwarfs in other polars
\citep{araujobetancoretal05-1}. Note that with $T_\mathrm{bb}=10^4$\,K
and $T_\mathrm{wd}=2\times 10^4$\,K, $F_\mathrm{in}$ for $x=1$ and
$x=100$ amounts to 16 times and 1600 times the nominal blackbody flux,
respectively.

% ---------- Fig. 2 -------------------------------------------------------
\begin{figure}[t]
\includegraphics[bb=39 62 584 774,angle=270,width=\columnwidth]{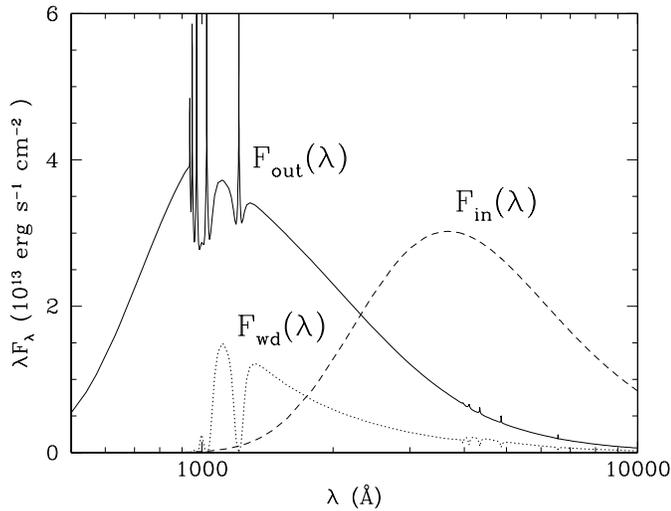}
\caption[ ]{Example for the relation between intrinsic spectral flux
$F_{\rm wd}$ of the unheated atmosphere of the white dwarf (dotted
curve), the low-temperature irradiation $F_{\rm in}$ (dashed curve),
and the outgoing flux $F_{\rm out}$ of the heated atmosphere (solid
curve).}
\label{fig-2}
\end{figure}
% -------------------------------------------------------------------------

\subsection{Case 1: Isotropic irradiation of a plane-parallel atmosphere}
\label{isotropic}

In Figure~\ref{fig-2}, we show an example of the relation between the
incoming and outgoing flux for $x = 4.6$, i.e., the outgoing flux
$F_\mathrm{out} = (1+x)\,F_\mathrm{wd} = 5.6\,F_{\rm wd}$ and the
effective temperature of the heated surface element is
$T_\mathrm{eff}= (1+x)^{1/4}\,T_\mathrm{wd}=3.08\times 10^4$\,K. The
optical/infrared irradiation is reprocessed into the outgoing
ultraviolet flux sufficiently high up in the atmosphere to cause an
inversion in the temperature stratification and the formation of emission
line cores.

\begin{figure*}[t]
\includegraphics[bb=39 43 589 769,angle=270,width=85mm]{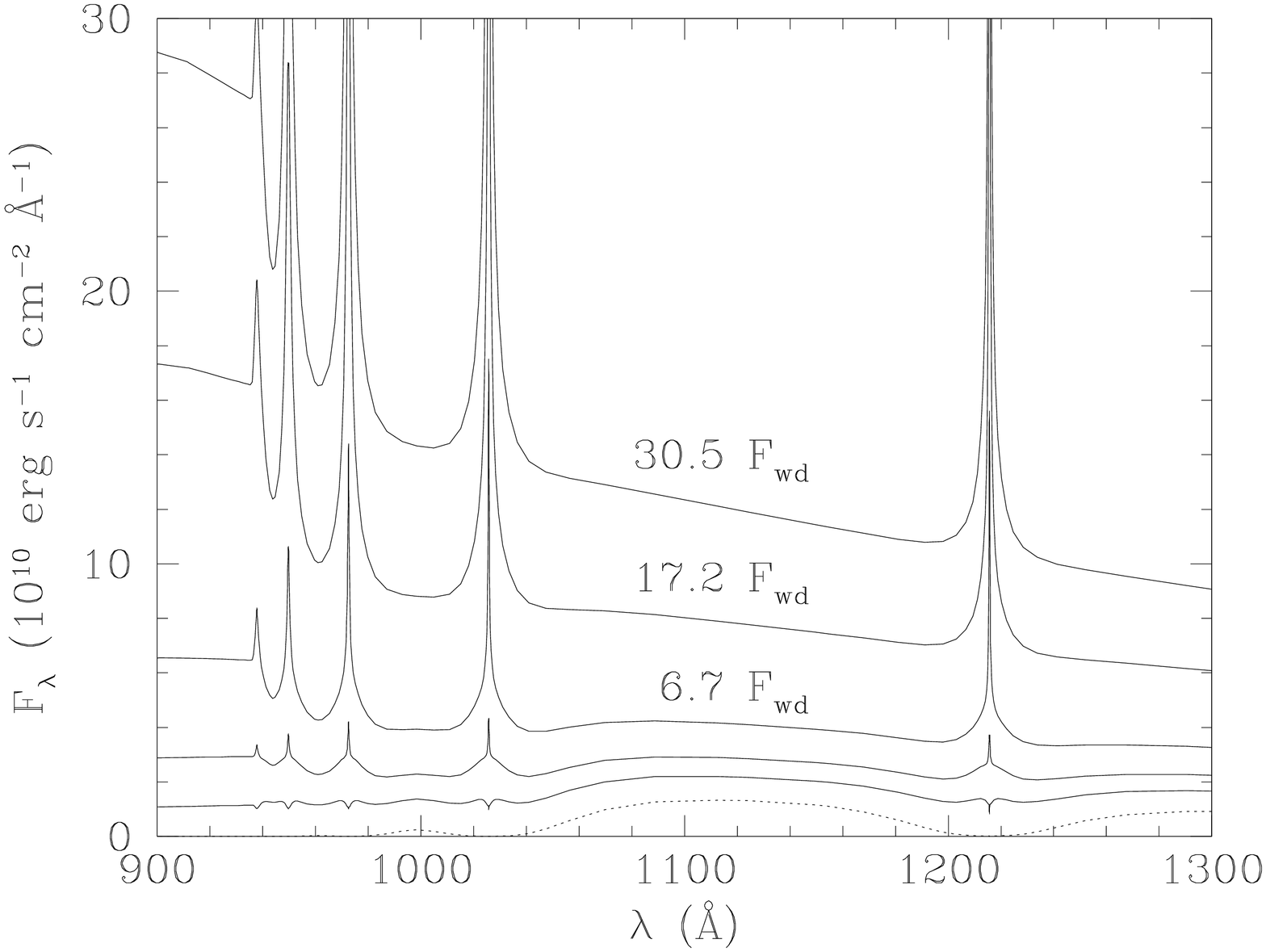}
\hspace*{5mm}
\includegraphics[bb=39 28 585 742,angle=270,width=85mm]{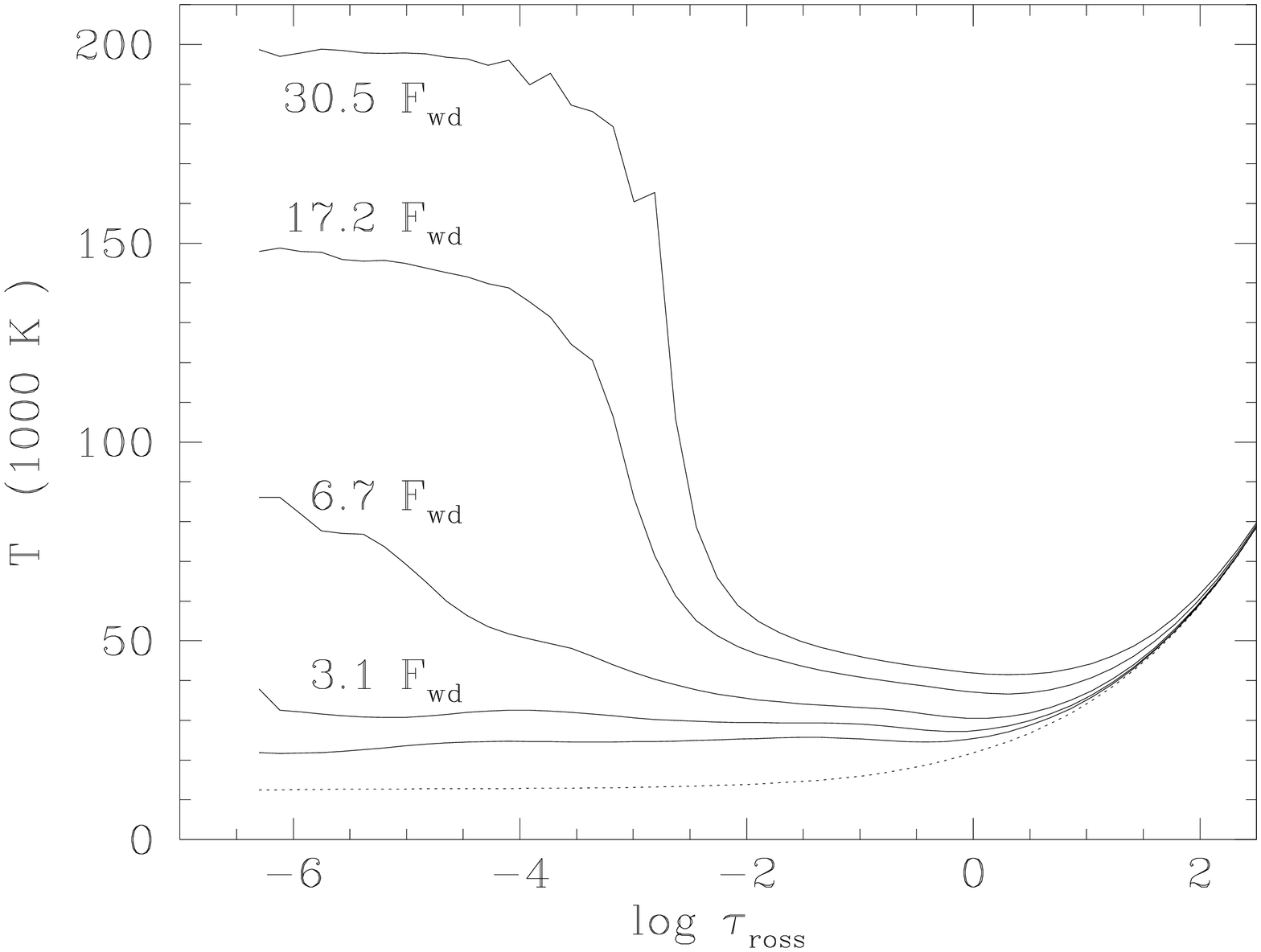}
\caption[ ]{Emerging spectral flux (left) and temperature
stratification (right) for atmospheres isotropically irradiated with
five different levels of the $10^4$\,K blackbody-like spectral flux
with 1.8, 3.1, 6.7, 17.2, and $30.5$ times the integrated flux $F_{\rm
wd}$ of the $2\times 10^4$\,K white dwarf atmosphere (solid curves, from bottom
to top). Also shown is the spectral flux (left) and the temperature
stratification (right) of the un-irradiated white dwarf (dotted
curves).}
\label{fig-3}
\end{figure*}

In Figure~\ref{fig-3}, we show the emitted spectral flux and the
temperature stratification for five selected values of $F_\mathrm{in}
= x\,F_{\rm wd}$ with $x = 1.8, 3.1, 6.7, 17.2$ and 30.5. Our
calculations cover a range of $x = 0.1$ to $\sim200$, with a model
for a lower value of $x$ acting as a start model for the next
higher $x$. For large $x$, instabilities in the temperature
stratification develop (see e.g. the model for $x = 30.5$ in
Fig.~\ref{fig-3}, right panel), which ultimately limit the manageable
irradiation level. Since we are still far from the Eddington limit, we
think that this deficiency can be mended in future applications.

For moderate irradiation with $x\la 3$, the temperature inversion
(Fig.~\ref{fig-3}, right panel) is barely noticeable and the
temperature stays essentially constant as a function of the Rosseland
optical depth $\tau_{\rm ross}$ until it merges into the temperature
profile of the unheated white dwarf at large optical depths.  For
$x\ga 3$, a pronounced temperature inversion evolves and for still
higher values, $x\ga 10$, a hot corona develops at optical depths
$\tau_{\rm ross}\la 10^{-2}$. As a consequence, the outgoing spectra
display filled-up Lyman lines for $x\la 3$ and emission cores
beyond. The toy model devised by \citet{gaensickeetal98-2} for AM~Her
assumed a temperature independent of $\tau_{\rm ross}$ in the outer
atmosphere and yielded filled up lines very similar to the situation
obtained here for $x\simeq2-3$.

% ---------- Fig. 3ab -------------------------------------------------------
% -------------------------------------------------------------------------

% ---------- Fig. 4ab -------------------------------------------------------
\begin{figure*}[t]
\includegraphics[bb=39 56 590 769,angle=270,width=87mm]{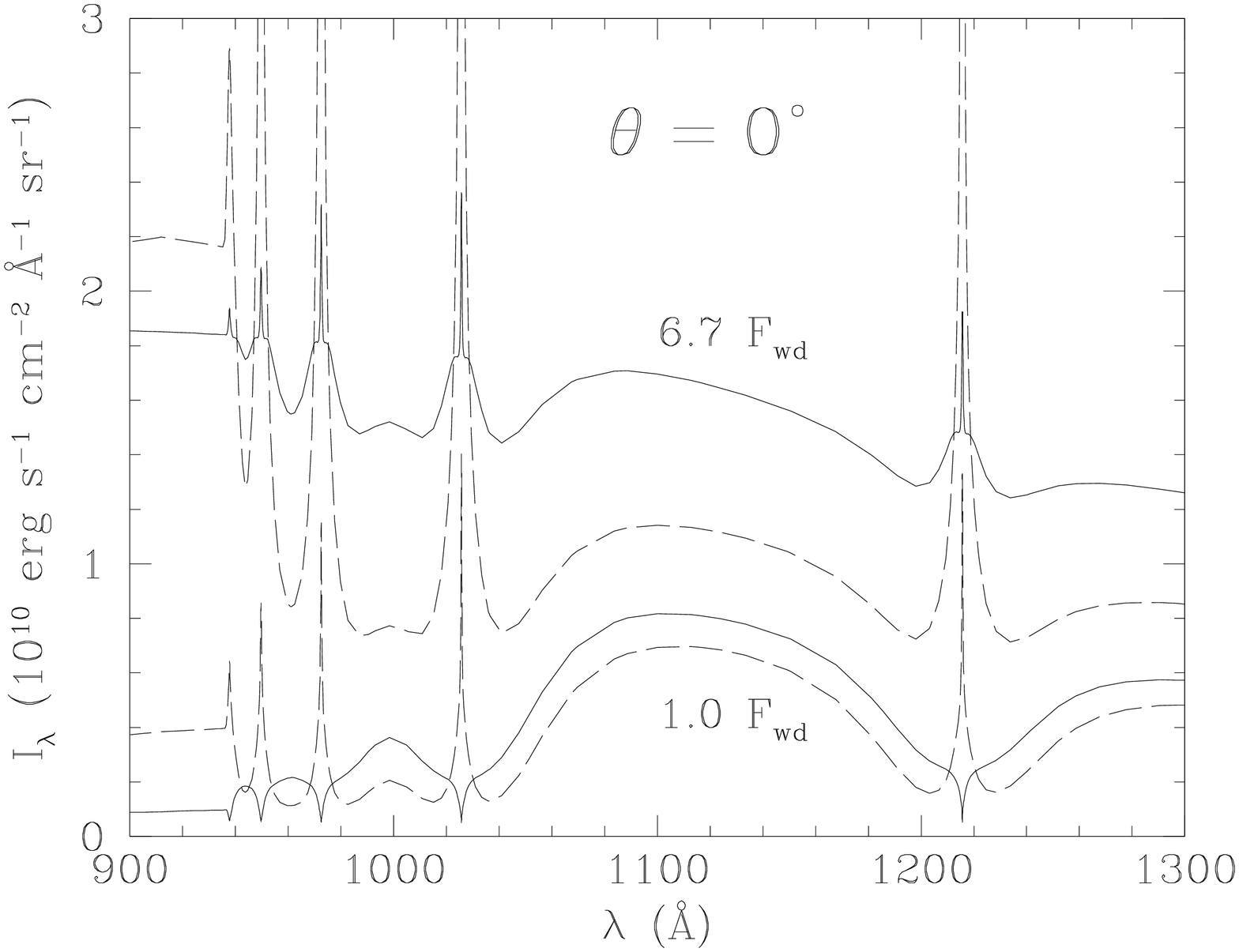}
\hfill
\includegraphics[bb=39 56 590 769,angle=270,width=87mm]{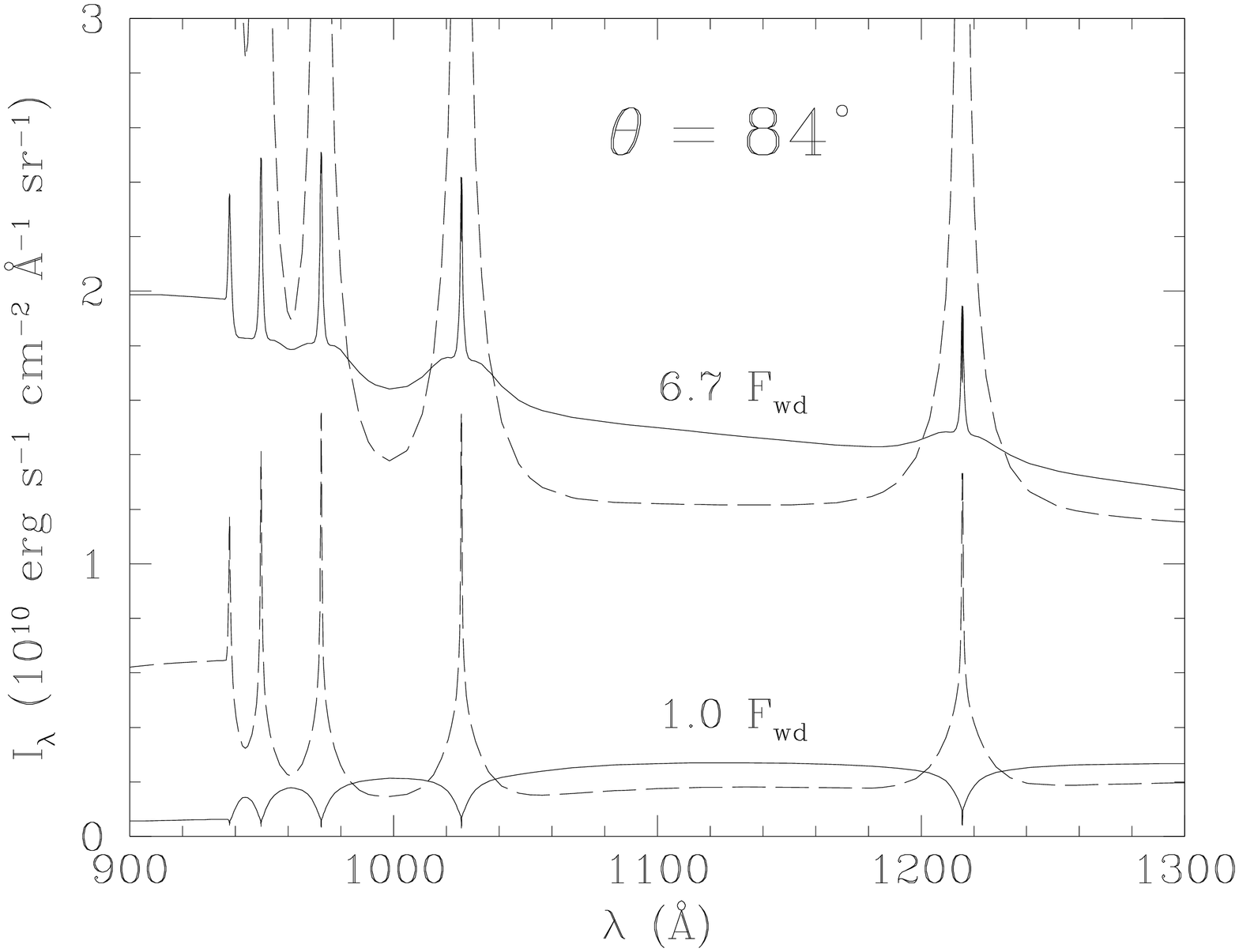}
\caption[ ]{Directional fluxes at emission angles $\theta=0^\circ$
(looking perpendicular into the irradiated atmosphere, left panel) and
$\theta=84^\circ$ (looking under a grazing angle into the atmosphere,
right panel) relative to the radial direction for irradiation at
$\mu=1.0$ ($\vartheta=0^\circ$, solid curves) and $\mu=0.1$
($\vartheta=84^\circ$, dashed curves), shown for two levels of the
incident flux, $1.0\,F_{\rm wd}$ and $6.7\,F_{\rm wd}$ (lower and
upper set of curves).}
\label{fig-4}
\end{figure*}
% -------------------------------------------------------------------------

% ---------- Fig. 5ab -------------------------------------------------------
\begin{figure*}[t]
\includegraphics[bb=110 235 532 680,width=3.5cm]{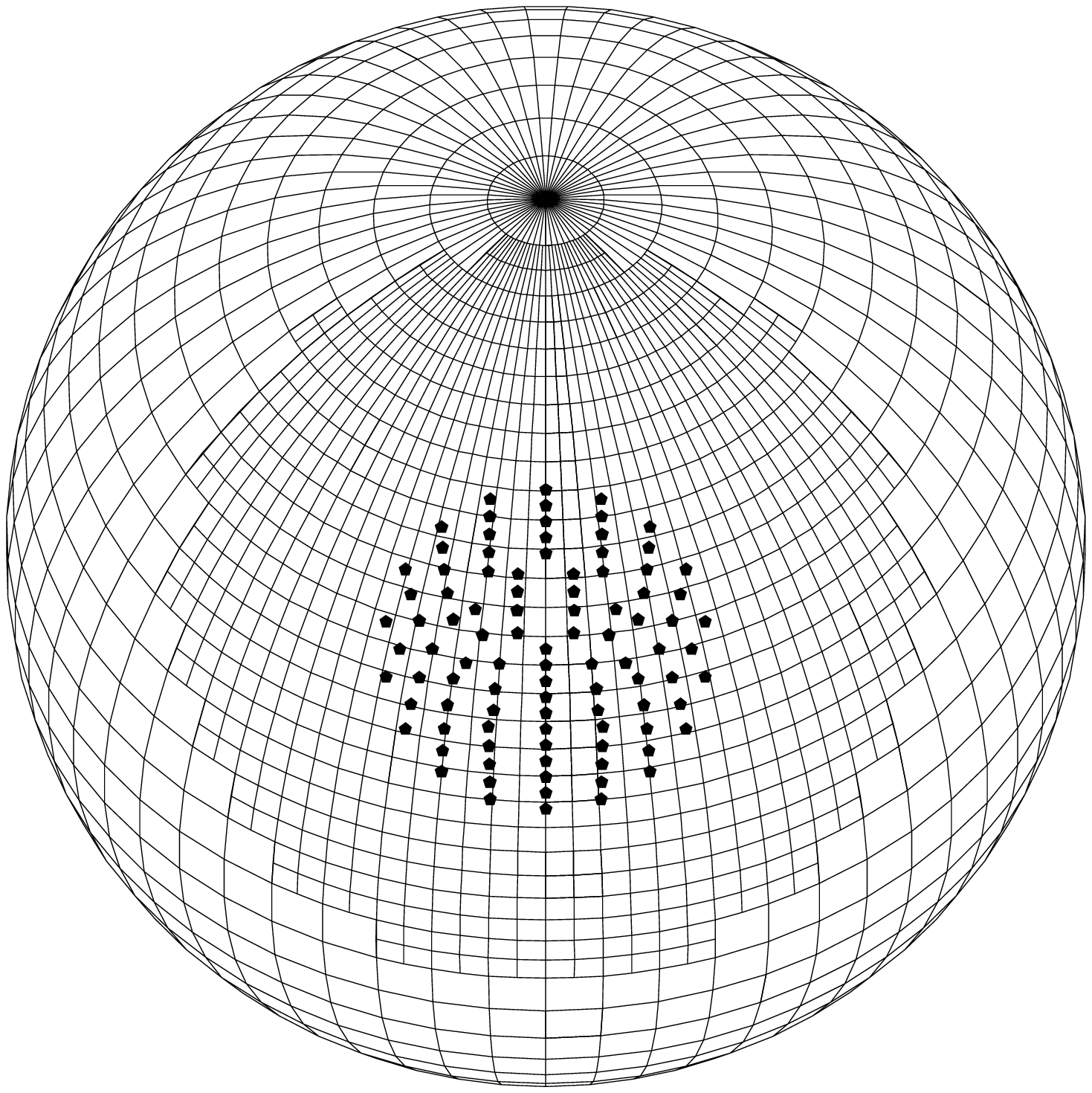}
\hspace{7mm}
\includegraphics[bb=110 235 532 680,width=3.5cm]{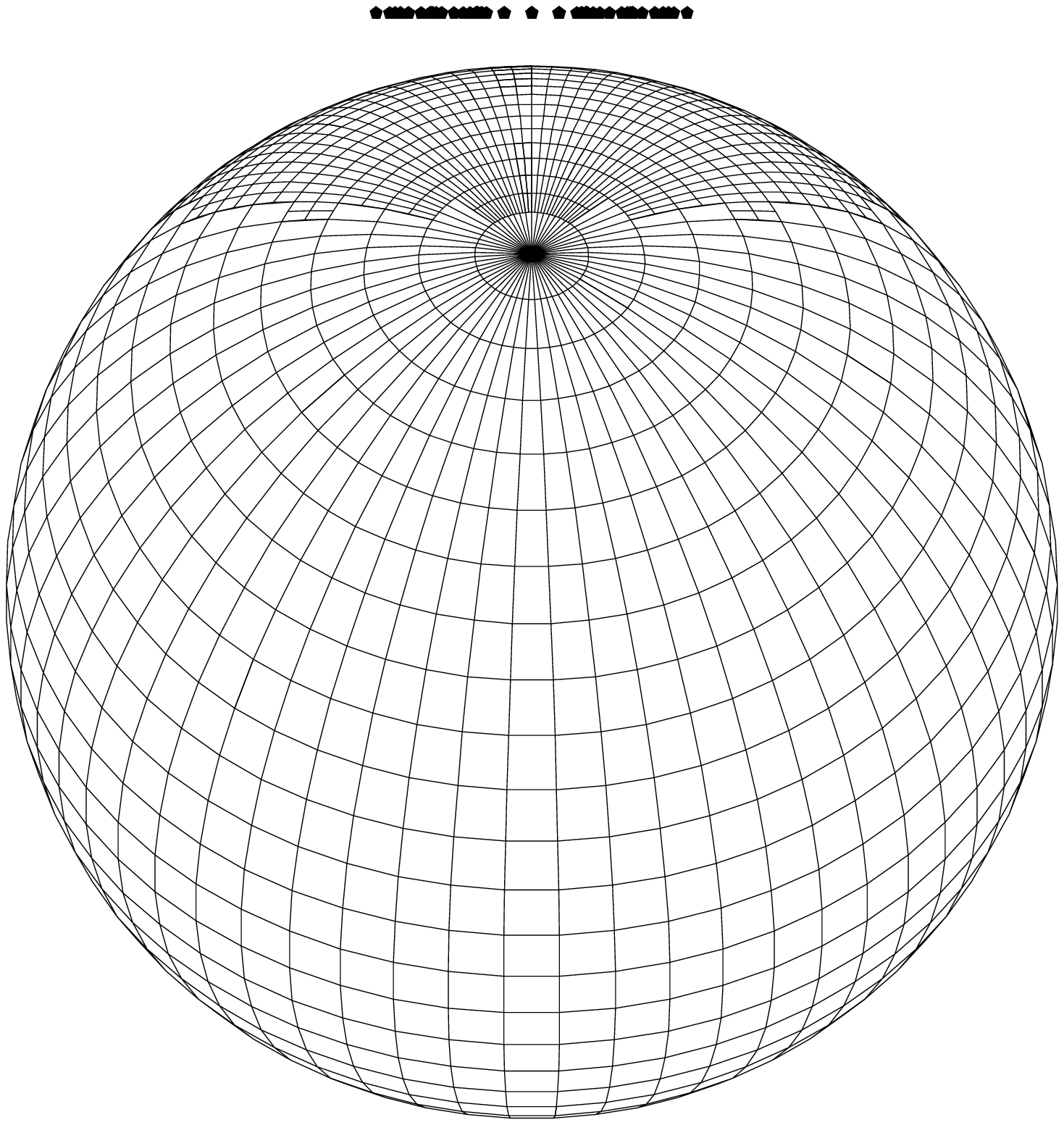}
\hspace{12mm}
\includegraphics[width=3cm]{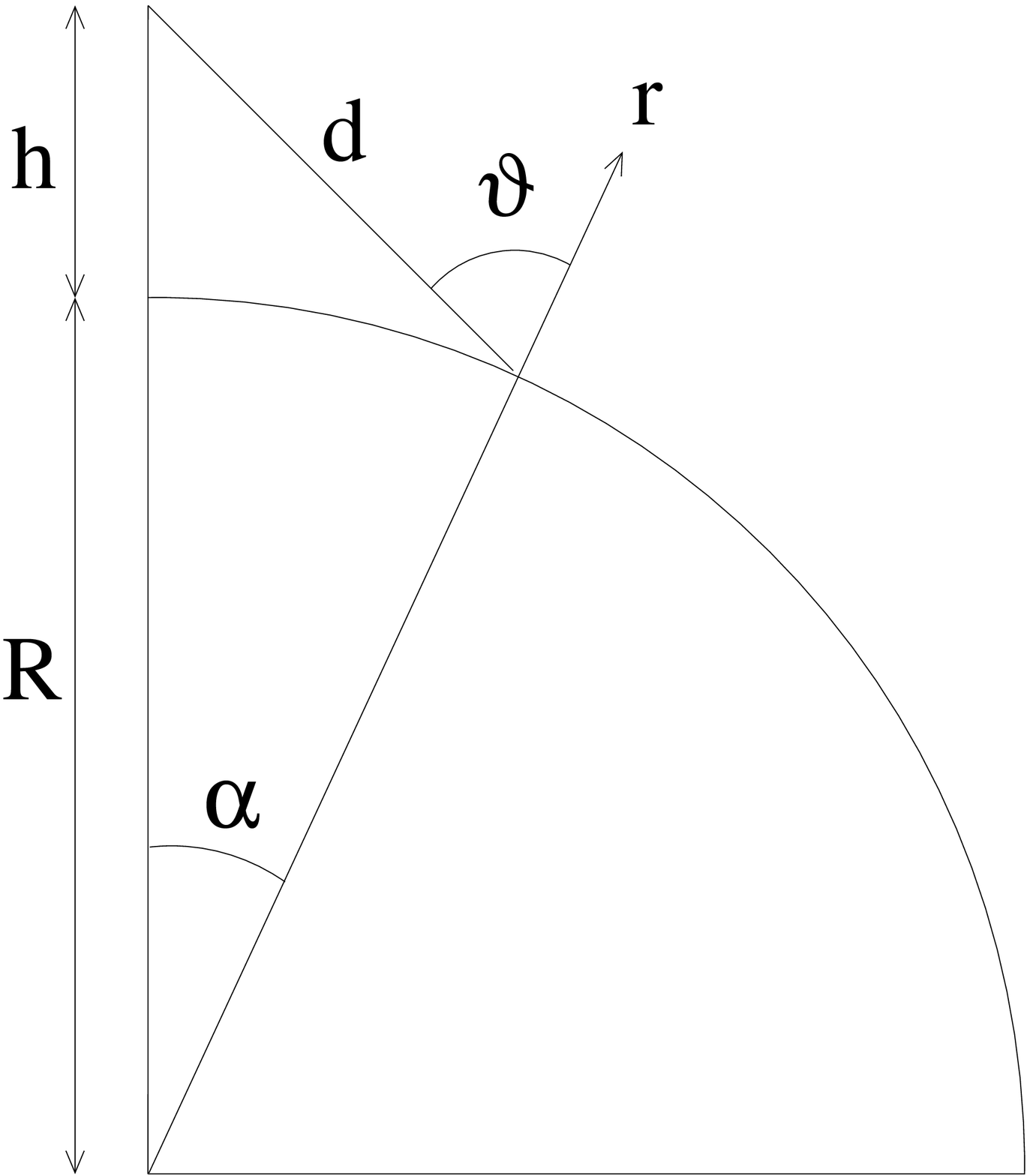}
\hspace{3mm}
\includegraphics[width=5cm]{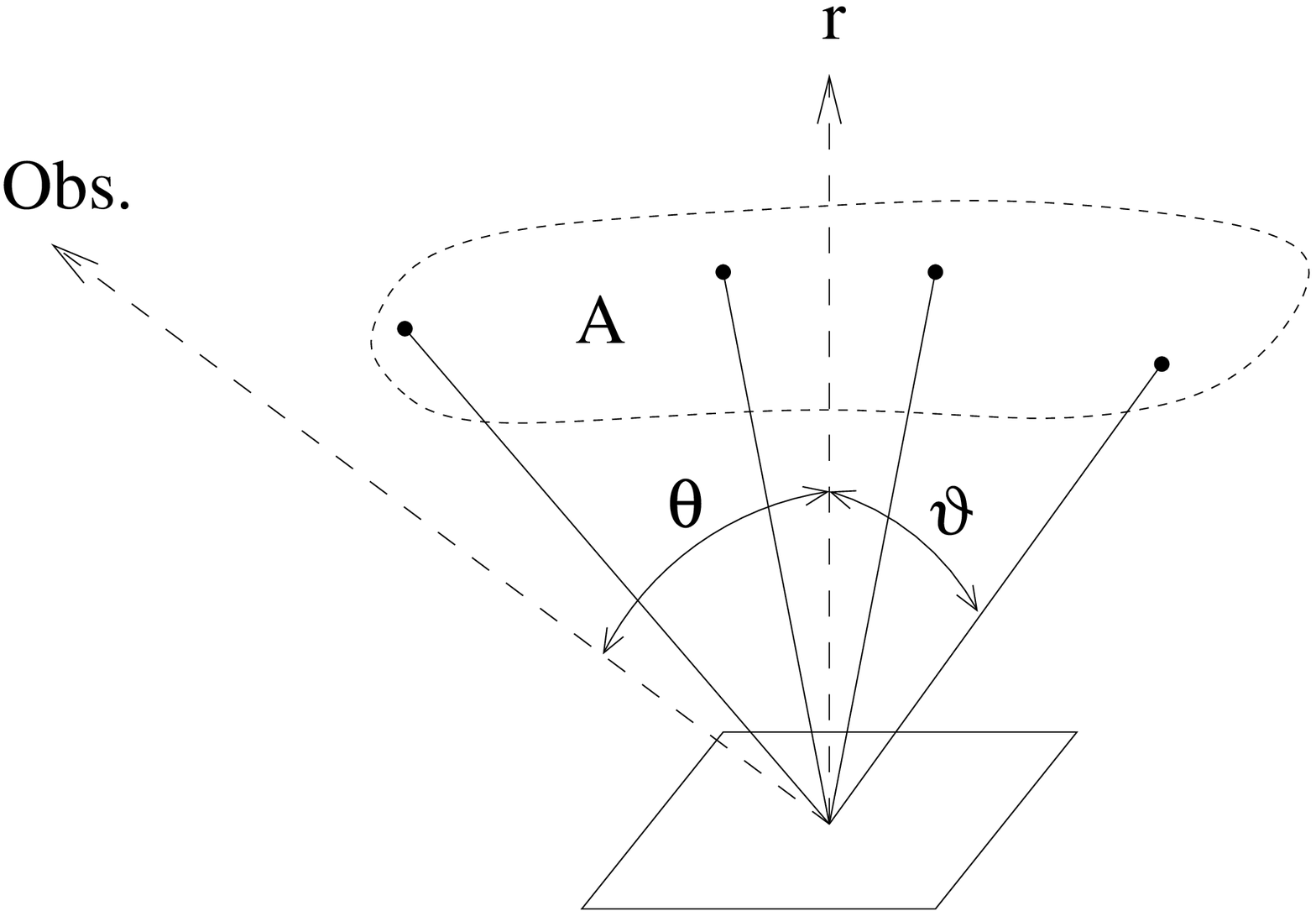}
\caption[ ]{\emph{Left: } Array of surface elements used in
calculating models of white dwarfs irradiated by an extended source at
a height $h=0.1\,R_{\rm wd}$ above the photosphere, shown for phases
$\phi=0$ (left) and $\phi=0.5$ (right).  \emph{Center: } Definition of
angles $\alpha$ and $\vartheta$.  \emph{Right: } Irradiation of a
single surface element by an extended source of area $A$ at height $h$
above the photosphere. $\theta$ indicates the direction to the
observer. }
\label{fig-5}
\end{figure*}
% -------------------------------------------------------------------------

\subsection{Case 2: Irradiation of a plane-parallel atmosphere at a fixed angle}
\label{angledependent}

We have calculated a 2-D grid of model atmospheres for the same range
of incident fluxes as above, $x=0.1$ to 200, and for direction cosines
$\mu = {\rm cos}\,\vartheta = 0.1$ to $1.0$ in steps of 0.1, with
$\vartheta$ the angle under which the irradiation occurs. For each
atmosphere with irradiation parameters $x, \mu$, our database contains
the intensities $I_\mathrm{\lambda}(\theta,x,\mu)$ as a function of
wavelength for a range of emission angles $\theta$, with
cos\,$\theta=0.1$ to 1.0 in steps of 0.1. Selected results for $x=1.0$
and 6.7 and $\mu=0.1$ and 1.0 are shown in Fig.~\ref{fig-4}. The left
panel is for $\theta=0^\circ$ (looking perpendicular into the
irradiated atmosphere), the right one for $\theta=84^\circ$ (looking
under a grazing angle into the atmosphere). In both panels, the solid
curves are for vertical irradiation ($\vartheta=0^\circ, \mu=1.0$) and
the dashed curves for grazing irradiation ($\vartheta=84^\circ,
\mu=0.1$). Finally, in both panels, the lower and upper pair of curves
are for irradiation parameters~ $x = 1.0$ and 6.7, respectively. The
individual curves show that irradiation incident vertically on the
atmosphere leads to weaker emission cores than grazing irradiation,
which is reprocessed high up in the atmosphere. Radiation leaving the
atmosphere vertically originates at larger depths and displays broader
absorption lines compared with the radiation leaving the atmosphere at
\mbox{large $\theta$}. Low-level irradiation ($x=1$) preserves the
limb darkening of the un-irradiated atmosphere, whereas grazing
irradiation at the higher irradiation level ($x=6.7$) leads to limb
brightening in the continuum and even more so in the lines. While this
is what one might expect intuitively, the quantitative results allow
us to construct models of accretion-induced irradiation for more
complicated geometries.

\subsection{External point source and extended source}

We consider two models of the radiation received from the entire white
dwarf irradiated by a source of luminosity $L$. The simpler model
assumes a point source located at a height $h$ above the surface
and the more involved one assumes an extended source at the same
height, consisting of an array of $N$ point sources of equal
luminosity $L_{\rm i}=L/N$, with $i = 1$ to $N$. Fig.~\ref{fig-5}
illustrates the irradiation geometry.

For irradiation by a point source, the spectral flux
$f_\mathrm{\lambda}$ observed at Earth is given by the sum over
the contributions from the $m$ surface elements visible for a given
orientation of the spotted white dwarf from the direction of the
observer,
\begin{equation}
f_\mathrm{\lambda} = \frac{1}{D^2}\sum_{j=1}^m A_j{\rm
cos}\,\theta_j\,I_\mathrm{\lambda}(\theta_j,
x_j,\mu_j).
\end{equation}
Here, $D$ is the distance between the observer and the source,
$A_j$ and cos\,$\theta_j$ the area and the cosine of
the viewing angle of the \mbox{$j$-th} surface element with
$\theta_j$ the angle between the normal to this element and
the direction to the observer, and $x_j$ and $\mu_j$
the irradiation parameters of the element as defined above, of which
$x_j$ is given by
\begin{equation}
x_j=\frac{F_\mathrm{in,j}}{\,F_\mathrm{wd}}=\frac{1}{\,F_\mathrm{wd}}\,\frac{L}{4\pi
d_j^{\,2}}\, \mu_j.
\end{equation}
with $d_j$ the distance between the irradiation source and
the \mbox{$j$-th} surface element as depicted in the central graph in
Fig.~5.  Since each surface element is characterized by uniquely
defined values of $x_j$ and $\mu_j$,
$f_\mathrm{\lambda}$ can readily be calculated using the spectral
intensities $I_\mathrm{\lambda}(\theta_j,
x_j,\mu_j)$ from the database.

% ---------- Fig. 6 -------------------------------------------------------
\begin{figure}[t]
\includegraphics[bb=40 33 583 753,angle=270,width=\columnwidth]{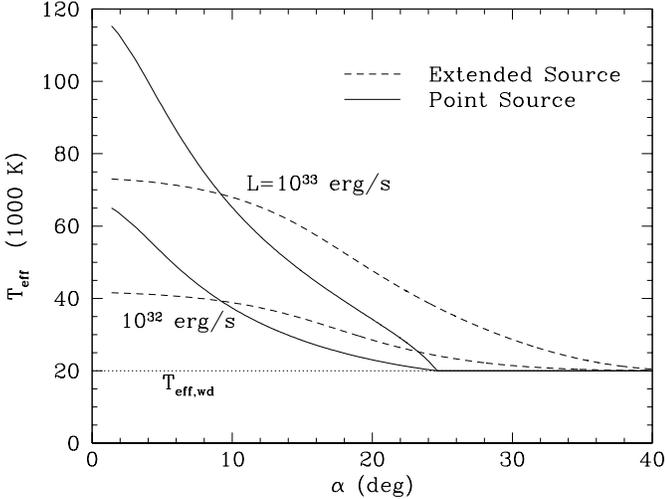}
\caption[]{\label{fig-6} Effective temperature in the large
irradiation-heated spot on a 0.6\,\Msun\ white dwarf as a function of
the angular distance $\alpha$ of the irradiated surface element from
the center of the source. Irradiation by a point source (solid
curves) or an extended source of $0.3\,R_{\rm wd}$ radius with $R_{\rm
wd}=8.6\times 10^8$\,cm (dashed curves) was considered, both located
at a height of $0.1\,R_{\rm wd}$. The luminosity is $10^{32}$ for the
lower curves and $10^{33}$\,\ergs\ for the upper curves. The
horizontal dotted curve is for the unheated white dwarf. }
\end{figure}
% -------------------------------------------------------------------------

% ---------- Fig. 7 -------------------------------------------------------
\begin{figure}[t]
\centerline{\includegraphics[bb=40 63 584 742,angle=270,width=84mm]{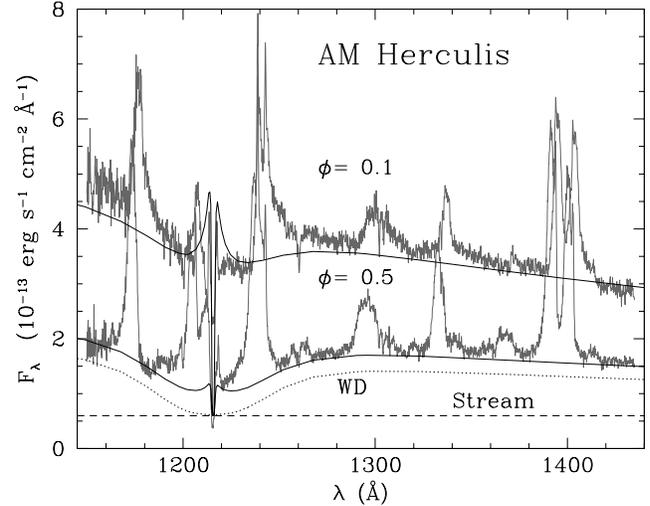}}
\caption[]{\label{fig-7} {\it HST/GHRS} spectrum of AM~Her in the
January 1997 high state at orbital phases $\phi=0.1$ (heated polar cap
in front) and $\phi=0.5$ (heated polar cap at horizon). The model
spectra are for irradiation by an isotropically emitting source of
$7.7\times 10^{32}$\,\ergs\ with radius $0.3\,R_{\rm wd}$ located at a height
$0.1\,R_{\rm wd}$ and for a distance of AM~Her of $D=79$\,pc. The
contribution by the accretion stream (dashed curve) and the spectral
flux of the unheated white dwarf (dotted curve, added to the dashed
curve) are included for comparison.  For the definition of the orbital
phase see \citet{gaensickeetal98-2}.}
\end{figure}
% -------------------------------------------------------------------------

The situation differs for an extended source as depicted in
Fig.~5. The right-hand graph shows a schematic and the left-hand
graphs the actual model in which the extended source is approximated
by a set of $N$ subsources of luminosity $L_{\rm i}=L/N$ with
$N=100$. Because of the curvature of the white dwarf surface, only a
subset $n_j\le N$ contributes to the irradiation of the \mbox{$j$-th}
surface element. The element is irradiated by each subsource with
index $i$ under its own angle of incidence $\vartheta_i$ with
$\mu_i=\mathrm{cos}\,\vartheta_i$. The irradiation parameter $x_j$ for
the $j$-th element is given by the first summation term
\begin{equation}
x_j=\frac{F_\mathrm{in,\,j}}{\,F_\mathrm{wd}}
=\frac{1}{\,F_\mathrm{wd}}\,\sum_{i=1}^{n_j}\frac{L_i}{4\pi d_i^2}\mu_i
=\frac{\langle \mu \rangle_j}{\,F_\mathrm{wd}}\,\sum_{i=1}^{n_j}
\frac{L_i}{4\pi d_i^2}.
\end{equation}
A complication arises from the fact that the $\mu_i$ in the sum cover
a finite range preventing the direct use of the database spectra. The
exact solution would require us to calculate for each single
surface element a custom-tailored atmosphere model taking into account
the irradiation fluxes from all subsources visible to that surface
element, which would be extremely time consuming. For simplicity, we
have opted, therefore, to replace the actual distribution of $\mu_i$
for the $j$-th surface element by a flux weighted mean
$\langle\mu\rangle_j$, as defined by the right-hand equality in
Eq.~(18), and use the database spectra $I_\mathrm{\lambda}(\theta_j,
x_j,\langle\mu\rangle_j)$ to calculate the observed spectral flux as
\begin{equation}
f_\mathrm{\lambda} = \frac{1}{D^2}\sum_{j\,=1}^m A_j{\rm
cos}\,\theta_j\,I_\mathrm{\lambda}(\theta_j,
x_j,\langle\mu\rangle_j).
\end{equation}

The radial temperature profiles of the spot created by irradiation
from a point source and an extended source differ substantially. As an
example, we compare in Fig.~\ref{fig-6} the temperature distributions
over the polar cap of a white dwarf irradiated by an isotropically
emitting point source of luminosity $L$ at a height
$h=0.1\,R_\mathrm{wd}$ (solid curves) and by an extended source of the
same total luminosity represented by an array of $N=100$ subsources at
the same height and distributed over an area with a radius
$r=0.3\,R_{\rm wd}$ (dashed curves). We use $R_\mathrm{wd}=8.6\times
10^8$\,cm appropriate to a 0.6\,\Msun\ white dwarf.  The upper and
lower set of curves are for $L=10^{33}$\,\ergs\ and
$L=10^{32}$\,\ergs, respectively. The abscissa is the angular
separation $\alpha$ from the center of the source, as defined in
Fig.~\ref{fig-5} (center graph).
The extended source is seen to heat a larger area to a lower peak
temperature. These temperature distributions do not account for
cyclotron beaming, which would broaden them, nor for contributions
from the stream or the X-ray source, which would narrow them.

\section{Application to HST/GHRS observations of AM~Her}

AM~Her, the bright prototype of polars, oscillates irregularly between
states of high and low (or possibly zero) accretion on time scales of
months to years and represents, therefore, an ideal laboratory in
which to study the irradiation-heated white dwarf. Since the discovery
that the far ultraviolet continuum of AM~Her is dominated by the white
dwarf \citep{heiseverbunt88}, numerous spectroscopic studies have
added to our knowledge of the response of its atmosphere to accretion
and accretion-induced irradiation. In the low state, a deep \La\
absorption line is observed at all orbital ($\equiv$ rotational)
phases, which becomes largely filled-up in the high state
\citep{gaensickeetal95-1, gaensickeetal98-2, gaensickeetal05-1}. The
same holds for the higher Lyman lines \citep{maucheraymond98}. The
lack of absorption is illustrated by the conclusion of
\citet{greeleyetal99} that the continuum is surprisingly well
represented by a blackbody. Our calculations add to an understanding
of the atmospheric structure of the white dwarf in AM~Her and provide
an internally consistent explanation for the lack of deep Lyman
absorption lines. Since AM~Her is typical of the class of polars, they
add to our understanding of polars, in general.

For our specific model calculations, we use a distance of AM~Her
$D=79$\,pc \citep{thorstensen03}, a radius of the white dwarf
$R_\mathrm{wd}=8.6\times 10^8$\,cm appropriate for a mass of
0.6\,\Msun\ \citep{gaensickeetal95-1}, and an inclination
$i=40^\circ$, which is within the range of $i = 35^\circ - 60^\circ$
reported for AM~Her \citep{gaensickeetal01-1}. We add interstellar
absorption in the cores of the Lyman lines using an interstellar
column density of atomic hydrogen, $N_{\rm H}=(3\pm1.5)\times
10^{19}$\,H-atoms\,cm$^{-2}$, which reproduces the width of the narrow
absorption core observed in the \La\ line of AM~Her
\citep{gaensickeetal98-2}. The continuum spectral shape is not
affected noticeably by the observational uncertainty in the column
density. The height and radius of the irradiation source are taken as
$h=0.1\,R_\mathrm{wd}$ and $r=0.3\,R_\mathrm{wd}$, respectively, which
yield the observed spot size of $\sim10$\% of the white dwarf surface
area \citep{gaensickeetal95-1, maucheraymond98,
gaensickeetal05-1}. With these assumptions, we calculated the observed
spectrum $f_\mathrm{\lambda}$ as a function of orbital phase
$\phi$. The two left-hand graphs in Fig.~5 illustrate the aspect of
the white dwarf at two orbital phases, $\phi =0$ (best visibility of
the spot) and $\phi =0.5$ (spot near the horizon, source located above
the white dwarf).

Fig.~7 shows the January 1997 high state Hubble Space Telescope {\it
GHRS} spectra of AM~Her at orbital phases $\phi=0.1$ and $\phi=0.5$
from \citet{gaensickeetal98-2} as grey curves (compare the
illustrations in Fig.~5). The spectra display strong Doppler-broadened
asymmetric emission lines of C{\scriptsize III}$\lambda 1176$,
Si{\scriptsize III}$\lambda 1207$, \La$\lambda 1216$, N{\scriptsize
V}$\lambda 1239,1243$, Si{\scriptsize III}$\lambda1299,1303$,
C{\scriptsize II}/C{\scriptsize II}$^*\lambda 1335$, and
Si{\scriptsize IV}$\lambda 1394,1403$, which originate in the
accretion stream. We estimate the continuum flux associated with the
stream emission as follows. \citet{gaensickeetal05-1} have compared
the $FUSE$ spectra of AM~Her at different accretion levels and find a
wavelength-independent stream contribution to the continuum flux for
the June 2000 high state of $0.4\times 10^{-13}$\,\ergcsa.  Converted
to the brighter January 1997 high state, this contribution is about
$0.6\times 10^{-13}$\,\ergcsa. \citet[ their
Fig.~1]{priedhorskyetal78} observed this component with U-band and
V-band fluxes of 0.6 and $0.2\times 10^{-13}$\,\ergcsa, respectively,
and noted that it decreases in the infrared as $1/\lambda^2$ . Fig.~7
shows the calculated spectra at $\phi=0.1$ and $\phi=0.5$ with the
stream contribution (dashed line) added. The resulting model spectra
are seen to provide an adequate fit to the observed continuum specta
at both orbital phases simultaneously (Fig.~7, solid curves). The eye
fit fixes the fractional luminosity intercepted and reemitted by the
extended spot to $2.3\times 10^{32}$\,\ergs, of the order of 10\% of
the bolometric luminosity. At a height of 0.1\,$R_\mathrm{wd}$,
the white dwarf subtends a solid angle of $1.2\pi$ and the
implied total luminosity of the irradiation source would be $7.7\times
10^{32}$\,\ergs, but this number may be misleading because it depends
on the actual height of the source and its angular emission
characteristics. Presently, we prefer simply to compare the integrated
model flux for $\phi=0.1$ at Earth with the observed fluxes of the
potential irradiation sources, which contribute to the heating of the
polar cap. The model flux longward of 912\,\AA\ at $\phi =0$ is
$3.3\times 10^{-10}$\,\ergcs, but an additional $6.0\times
10^{-10}$\,\ergcs\ or 65\% of the total is hidden in the Lyman
continuum. This large fraction of the Lyman continuum flux has not so
far been considered in the energy balace of AM~Her, but its existence
is supported by the observed pronounced Balmer and Paschen jumps in
emission in AM~Her's high state \citep{schachteretal91}. The fluxes of
the irradiation sources at Earth appropriate for the January 1997 high
state include a cyclotron flux of $2.0\times 10^{-10}$\,\ergcs\
\citep[ see also G\"ansicke et al. 1995]{baileyetal88}, a stream
contribution for $\lambda > 912$\,\AA\ of $3.3\times
10^{-10}$\,\ergcs\ plus probably a substantial Lyman continuum flux,
and a hard X-ray flux converted from the April 1991 high state value
of $1.3\times 10^{-10}$\,\ergcs\ \citep{gaensickeetal95-1} to
$1.6\times 10^{-10}$\,\ergcs, assuming that the conversion factor is
the same as in the FUV. The cyclotron flux complemented by fractions
of the stream and hard X-ray components can account for the observed
flux of the heated polar cap longward of 912\,\AA. The predicted Lyman
continuum flux from the polar cap requires an additional
source. Candidates are the Lyman continuum emission from the lower
section of the optically translucent accretion stream and the
potential soft X-ray emission from a spray of accreted matter
lifted above the photosphere of the white dwarf as a reaction to
the drastic fluctuations in ram pressure which occur in non-stationary
accretion. It is clear that the hard X-ray source can not play a
major role because of its limited luminosity and the low conversion
efficiency of hard X-rays, as addressed in the next section.

\section{Discussion}

We have quantitatively reproduced the $1150-1450$\,\AA\ continuum
spectra of AM~Her in its high state of accretion by a model that
considers a large polar cap heated by irradiation from the accretion
region.  Irradiation strongly affects the temperature stratification
of the white dwarf atmosphere in the heated polar cap. Strong heating
leads to an inversion of the radial temperature profile in the
atmosphere, causes the Lyman absorption lines to fill up completely,
and produces substantial emission in the Lyman continuum. The known
irradiation sources can account for the observed reprocessed flux
longward of the Lyman limit, but an additonal source is needed to
account for the predicted Lyman continuum flux. The implied
luminosity of the polar cap is larger than considered previously
\citep{gaensickeetal95-1} and stresses the point that the reprocessed
flux emerges primarily in the FUV and not in the soft X-ray regime as
predicted by the early simple models of polars \citep{kinglasota79,
lambmasters79}.

Since X-ray heating \citep{vanteeselingetal94} has not been considered
here it is appropriate to discuss the potential contribution from that
source in some detail. The horizontal temperature profile of the spot
vs. distance from the spot center and the vertical atmospheric
temperature profile at a given position may both be expected to differ
for irradiation with either optical/infrared radiation or hard X-ray
bremsstrahlung. Cyclotron radiation will produce a larger spot than
bremsstrahlung because it originates higher up in the post-shock
region and its beaming properties allow it to reach regions further
away from the spot center. This difference is strenghtened by the
lower shock heights of the denser, bremsstrahlung-dominated shocks
(Eq.~14) and by the fact that bremsstrahlung is emitted isotropically
and irradiation is thereby concentrated closer to the spot
center.  For a given surface element, one can judge the efficiency of
heating by cyclotron radiation and by bremsstrahlung as follows. As
seen from Fig.~3 (right panel), cyclotron heating is most intense at
Rosseland optical depths $\tau_{\rm ross} \la 0.01$ but reaches down
to the photosphere and beyond. Hence, cyclotron radiation is
completely reprocessed fairly high in the atmosphere. Estimating the
efficiency of bremsstrahlung heating requires us to consider the
competition between photoabsorption and Compton scattering. Hydrogen
is 99\% ionized at the relevant levels of our atmospheres, which still
allows atomic transitions to determine $\tau_{\rm ross}$ and causes
the electron scattering optical depth to become $\tau_{\rm es} \ll
\tau_{\rm ross}$. For the typical 10\,keV bremsstrahlung photons, on
the other hand, photoabsorption is negligible with $\tau_{\rm ph} \ll
\tau_{\rm es}$. Hence, bremsstrahlung photons will penetrate far below
the photosphere and are preferentially (back)scattered rather than
photoabsorbed with the consequence of ineffective atmospheric heating
\citep{vanteeselingetal94}. Heating by bremsstrahlung may be
important, however, if a finite metalicity of the atmosphere
drastically increases the absorption cross section. Since the
metalicity of the atmosphere will be enhanced in the inner spot where
accretion occurs, X-ray heating may dominate there, while cyclotron
heating is expected to dominate further away from the spot
center. While a more detailed model should take both irradiation
sources into account, we presently lack information on parameters like
the mass inflow rate per unit area and the ensuing metalicity as
functions of distance from the spot center.

The heating of the extended polar cap from the post-shock source tends
to zero for vanishing shock height because then all downward directed
emission is intercepted by the atmosphere within the proper accretion
spot.  Disregarding irradiation by emission from the pre-shock stream
for the moment, our model predicts that high-field polars, in which
all cyclotron and X-ray emission occurs in low-lying shocks (see
Eq.~14), should lack strongly irradiation-heated polar caps. This is,
in fact, the case for the 200\,MG system AR~UMa in its low state with
some remnant cyclotron emission and an amplitude of the orbital
modulation at 1400\AA\ of only 5\% \citep{gaensickeetal01-1}.  An
investigation of the temperature distributions of a larger sample of
white dwarfs in polars would shed light on some poorly understood
aspects of accretion physics. Such a study would have to consider
cooling of the deeply heated polar cap as well as the relevent heating
processes. Note that accreting white dwarfs are generally hotter than
single white dwarfs of the same age \citep{araujobetancoretal05-1} as
a result of compressional heating \citep{townsleybildsten03}, but
this process is rooted in the entire non-degenerate envelope (and to
some extent in the core) and heats the entire star and not just the
polar cap where accretion takes place.

Finally, a source of uncertainty in the calculated spectra warrants
mentioning. Our atmospheric model disregards the compression of the
outer atmosphere by accretion and is applicable strictly only outside
the region of the extended source, assumed to be fed by accretion.
The bombardment of the atmosphere by charged particles adds an
external pressure and affects the structure of the upper atmosphere
where the emission lines originate. The physics of bombarded
atmospheres is complicated and has been treated to a certain extent by
\citet{woelkbeuermann92}. A simple estimate suggests that the effect
may not be negligible. For a mass flow rate $\dot m =
10^{-4}$\,g\,cm$^{-2}$s$^{-1}$, which is below present detection
techniques, the ram pressure $P_{\rm ram} = \dot m \upsilon_{\rm ff}$,
with $\upsilon_{\rm ff}$ the free-fall velocity, amounts to a few
percent of the photospheric pressure and severely compresses the
atmosphere only outside $\tau_{\rm ross} \simeq 3\times 10^{-4}$. A
mass flow rate ten times higher, however, reaches down to $\tau_{\rm
ross} \simeq 0.05$. Under these circumstances, a more detailed study
of the line profiles and the strength of the Lyman continuum in
irradiated atmospheres of mCVs, including NLTE effects, requires a
substantially larger effort.

\begin{acknowledgements}
MK was supported by the Deutsche Forschungsgemeinschaft DFG within the
Graduiertenkolleg ``Str\"omungsinstabilit\"aten und Turbulenz``. BTG
was supported by a PPARC Advanced Fellowship. We thank Sonja Schuh and
the anonymous referee for useful comments on the manuscript.
\end{acknowledgements}

%\bibliographystyle{aa}
%\bibliography{irradiation}

\end{document}